\documentstyle[aps,preprint,epsfig]{revtex}

\begin{document} 
\draft

\title{
On geometric potentials in nanomechanical circuits
      }

\author{Alexander V. Chaplik$^{1}$ and Robert H. Blick$^{2}$}

\address{
$^{1}$ Institute of Semiconductor Physics, RAS, Siberian Branch, 630090 Novosibirsk, Russia. \\
$^{2}$ Electrical and Computer Engineering, University of Wisconsin-Madison,
1415 Engineering Drive, WI 53706, USA. \\
}

\date{April-2, 2002}
\maketitle

\begin{abstract}
We demonstrate the formation of confinement potentials in suspended nano\-structures induced by the geometry
of the devices. 
We then propose a setup for measuring the resulting geometric phase change of electronic
wave functions in such a mechanical 
nanostructure. The device consists of a suspended loop through which a phase coherent current is driven. 
Combination of two and more geometrically induced potentials can be applied for creating mechanical
quantum bit states.
\end{abstract}

\pacs{03.65.Vf,85.35.Ds,85.85.+j}

Quantum mechanics in curved-linear manifolds has been elaborated for some time~\cite{one}. The
propagation of waves in curved waveguides can be 'translated' for quantum particles into 
a Hamiltonian consisting of
the kinetic energy operator and a
resulting potential energy, which is of pure geometric origin.
The ability to build nanostructures with a three-dimensional relief allows the realization of 
low-dimensional electronic systems possessing a mechanical degree of freedom~\cite{blick}. This is exemplified in
recent work by Prinz~{\it et al.}~\cite{prinz} and Schmidt and Eberl~\cite{schmidt}, who demonstrated
how to realize rolled-up semiconductor films with a radis of curvature $R \sim 100$~nm.
Thus it is worth while studying the influence of geometrical potentials on phase coherently propagating 
particles in curved low-dimensional electron systems. This will induce a phase shift in the electronic
wave function corresponding to Berry's phase~\cite{berry}.

For the case of a two-dimensional electron gas, 
flexing it leads to a geometrical potential of the form

\begin{equation}
          U = -\frac{\hbar^2}{8 m} \left( \frac{1}{R_1} - \frac{1}{R_2} \right)^2
\end{equation}

where $m$ is the effective mass and $R_1$, $R_2$ are the principal curvature radii of the surface in the 
point where the electron resides. The geometric potential is
always attractive and independent of the electric charge
of a particle, similar to gravitation. Furthermore, it is of purely
quantum origin, i.e. it vanishes for the limit $\hbar \rightarrow 0$. 

If one of the radii tends to infinity we
obtain a cylindrical surface. Particularly, this is the
case when electrons are confined to a quantum wire having 
the shape of a plane curve. The Schr\"odinger equation
for such a curved linear 1D system reads (see Fig.~1(a)) 

\begin{equation}
\hat{H} \psi = - \left[ \frac{\hbar^2}{2m} \frac{d^2}{ds^2} + \frac{\hbar^2}{8mR^2(s)} \right] \psi = E \psi,
\label{hamiltonian}
\end{equation}

where $s$ is the length of the arc of the wire counted
from an arbitrary origin and $1/R(s)$ is the local
curvature. A straight forward derivation of Eq.~(\ref{hamiltonian}) is given in~\cite{vedez}. 
It is further shown in this work that a wire of the shape
of an Archimedes spiral gives the geometric potential with
an asymptotic behavior resulting in Coulomb's law

\begin{equation}
1/R^2(s) \sim 1/s.
\end{equation}

Hence, there is an infinite set of bound states corresponding to the localization of electrons
at the origin of the spiral. 

If one starts from a quantum wire consisting of two straight lines conjugated by an arc of a circumference
('open book'-shape) the corresponding geometrical potential is a rectangular potential well 
(see Fig.~1(b)) with a width $\alpha R$ and a depth $-\hbar^2/8mR^2$. Here $R$ is the radius
of the circumference and $\alpha$ is the conjugating arc (angle between two rectilinear 
parts of the wire). There exists one and only one bound state for
$\alpha < \pi$ in such a system and its energy is given by

\begin{equation}
E_o = - \frac{\hbar^2}{8mR^2} \left( 1 - \frac{16 z^2}{\alpha^2} \right),
\end{equation}

where $z(\alpha)$ is the root of $\cos z = 4z/\alpha$ 
between 0 and  $\pi/2$ (see Fig.~1(b)). For example, for a U-shaped wire 
($\alpha = \pi$) with $R = 100$~\AA and an electron mass in GaAs 
$m = 0.07~m_0$, we find a binding energy of $E_o = 4$~K. Whereas for 
the conjugation of two perpendicular straight lines, i.e. $\alpha = \pi/2$,
we obtain $E_o = 3$~K.
The phase of the wave function in the quasiclassical
regime is then given by the integral

\begin{equation}
\frac{1}{\hbar} \int_0^S P ds' = \frac{1}{\hbar} \int_0^S \sqrt{2 m (E_F - U(s')) }ds'.
\label{path}
\end{equation}

For a wire with a small curvature the relation for the Fermi energy and the 
geometrical potential is $E_F \gg \hbar^2/8mR^2$. The 
total phase shift $\Delta \phi$ after passing through the conjugation 
then is $\Delta \phi \cong \alpha / 8 k_F R$ where $k_F R \gg 1$ with $k_F = P_F/\hbar$. 
In the opposite limit, i.e. $k_F R \ll 1$ the integral in Eq.~(\ref{path}) 
gives $\Delta \phi = \alpha/2$. 

The most realistic case of course is given in
the limit $k_F R \gg 1$. A mechanically deformable quantum interferometer will be able
to sense such a deformation, provided its sensitivity to the phase shifts
exceeds the value $\alpha / 8 k_F R$. Such a mechanical quantum interferometer (MQUI)
can be realized by suspending a two-dimensional electron gas in a thin membrane. While
the electron gas usually is 10~nm thin, the total membrane thickness will be around 90~nm.
This leaves the minimal radius of curvature at $R \sim 500$~nm, giving a confinement
potential for the lowest state of $E_o \sim 1.6$~mK. As Prinz~{\it et al.}~\cite{prinz}
have shown smaller curvature radii can be achieved. Especially, considering surface bound 
two-dimensional electron gases in InAs heterostructures should allow reaching the regime
$R \sim 10$~nm, effectively leading to temperatures in the range of $\sim 4$~K.

An MQUI is shown in Fig.~2: the interferometer basically consists of a ring shaped suspended membrane containing a 
two-dimensional electron gas. This geometry first allows to measure interference induced by
a magnetic field applied perpendicular to the plane of the ring, i.e. 'classical' 
Aharonov-Bohm oscillations (see Fig.~2(a)). In this way the interferometer is to be calibrated. The diameter of
the ring should be smaller than the phase coherence length $L_{\phi}$, i.e. in this case
5~$\mu$m as found for two-dimensional electron gases will be sufficient. 
Deformation of the arms of the ring is facilitated by using 
gating electrodes beneath the ring, also indicated in Fig.~2(b). 
In order to avoid depletion of the electron
gas the suspended heterostructure contains a highly n-doped GaAs back-layer. By gating the
interferometer arms individually two flexing modes can be supported. Both
modes are shown in Fig.~2(b). Since the potential shows a dependence
of $\sim R^{-2}$ both will lead to the same wave function shift. 
Another technique for mechanically modulating the suspended electron ring is given by using 
surface acoustic waves~\cite{wixforth}. The modulation amplitude is considerably increased
in suspended membranes~\cite{beil}. 

Intriguingly the combination of two curved sections connected by a thin wire~$w$ -- denoted
as a $\Pi$-shaped element -- will lead
to a double quantum well potential, as demonstrated in Fig.~3(a). Depending on $R(s)$ and on~$w$ the two discrete 
states in the wells $E_o^{A,B}$ can communicate, i.e.\ a tunnel splitting of the order of 
$2 \delta E_o$ will occur. Thus this 
system represents a mechanical quantum bit (mqubit), whose communication is given by the exchange 
of phonons at an energy $2 \delta E_o = h f_{\rm ph} = h c_{\rm a}/w$, where $c_{\rm a}$
is the velocity of sound in the heterostructure and $w$~is the length of the connecting element. 
Naturally, this scheme can be extended to a chain
with $N$ elements forming $N/2$ $\Pi$-wire mqubits as shown in Fig.~3(b). 
A whole variety of different modes is available for information exchange between two mqubits (ii). 
Stretching and contracting the wire elements individually allows to steer inter and intra-mqubit communication.
The overall information exchange is performed through low frequency 
phonon modes, e.g.\ indicated by dashed lines (iii). The energy of these modes is 
determined by $c_{\rm a}$ and the chain length. Sectioning into
sub-chains the interaction of the mqubits organized in a hierarchy.

As another example we imagine to illuminate the U-shaped wires for generating electron-hole
pairs forming excitons. Both electrons and holes and hence the excitons will be attracted to the bottom 
of the U-wires' geometric potentials. Since excitons do not obey the 
Pauli exclusion principle, the probability to capture more and more excitons will increase.


This work has been supported by the Deutsche Forschungsgemeinschaft (DFG) through 
the projects Bl/487-1\&\ 3 and the Russian Foundation for Basic Research via 
grant 16377, by the program 'Physics of Solid Nanostructures' of the Russian Ministry 
of Science, and by the Sonder\-forschungs\-bereich (SFB 348) of the DFG.
AVC thanks Jorg Kotthaus for the hospitality 
during the research visit to the Center for NanoScience.
RHB likes to thank Herbert Walther for discussing the importance of 
vibronic modes for quantum computation in atom chains.

\bibliographystyle{prsty}

\begin{thebibliography}{10}

\bibitem{one} C.T. da Costa, Phys. Rev. A {\bf 23}, 1982 (1981);
H. Jensen and H. Koppe, Annals of Phys. {\bf 46}, 586 (1971);
N. Ogawa, K. Fujii, and A. Kobushukin, Progr. Theor. Phys. {\bf 83},894 
(1990). 


\bibitem{blick} R.H. Blick, F.G. Monzon, W. Wegscheider, M. Bichler, F. Stern, and M.L. Roukes, 
Phys. Rev. B {\bf 62}, 17103 (2000);
J. Kirschbaum, E.M. H\"ohberger, R.H. Blick, W. Wegscheider, M. Bichler, 
Appl. Phys. Lett. {\bf 81}, 280 (2002).

\bibitem{prinz} V.Ya. Prinz, V.A. Seleznev, A.K. Gutakovsky, A.V. Chehovskiy,
V.V. Preobrazhenskii, M.A. Putyato, and T.A. Gavrilova, Physica E {\bf 6}, 828 (2000). 

\bibitem{schmidt} Oliver G. Schmidt and Karl Eberl, Nature {\bf 410}, 168 (2001). 

\bibitem{berry} Michael V. Berry, Proc. R. Soc. London Ser. A {\bf 392}, 45 (1982). 

\bibitem{vedez} A.I. Vedernikov and A.C. Chaplik, ZhETF {\bf 11}, 448 (2000).
In this specific case the kinetic energy operator has the conventional form of 
a 1D Schr\"odinger equation, obtained by replacing '$x$' by '$s$'.
A curve is not possessing an internal geometry in contrast to the 2D-case
and manifolds of higher dimension. 

\bibitem{wixforth} A. Wixforth, J.P. Kotthaus, and G. Weimann, Phys. Rev. Lett. {\bf 56}, 2104 (1986).

\bibitem{beil} F.W. Beil, A. Wixforth, and R.H. Blick, Physica E {\bf 13}, 473( 2002).

\end{thebibliography}

\newpage

Fig.~1: (a) Sketch of a single wire and the resulting geometrically induced confinement potentials. The 
parameter $\alpha$ gives the degree of bending and $E_o$ represents a bound electronic state in the
resulting potential. (b) For $\alpha = \pi$ and $\alpha = \pi/2$ square well potentials with different
binding energies are obtained. 
\\

Fig.~2: Mechanical quantum interferometer: (a) electrons propagate phase coherently through the ring similar
to an Aharonov-Bohm geometry.
Flexing the arms of the ring ($\alpha$, $\alpha'$) a geometrical potential is formed which leads to an effective phase
shift of the electronic wave function within the interferometer. (b) Two flexing modes can be distinguished (i) and
(ii) both of which are leading to an identical phase shift. 
\\

Fig.~3: (a) A $\Pi$-wire forming a double quantum well potential: the discrete states in the two wells
can interact depending on the connecting wire element length~$w$. In case of tunneling a mechanical
quantum bit (mqubit) is formed, i.e. the two discrete states are tunnel split by $\delta E_o$.
(b) Chain of $\Pi$-wire elements defining a circuit of ten coupled mechanical quantum bits (i).
Communication between two mqubits is achieved by a variety of local deformations of the wires (ii).
Parallel addressing of mqubit chains is possible through 
acoustic phonons propagating along the wire with the velocity of sound (iii) but at lower
frequencies (indicated by dashed lines).
\\

\end{document}